\documentclass{aa}  

\usepackage{graphicx}
\usepackage{txfonts}
\begin{document}

   \title{On the origin of very massive stars around NGC\,3603}

   \author{V. M. Kalari
          \inst{1,2}, J. S. Vink\inst{3}, W. J. de Wit\inst{4}, N. J. Bastian\inst{5} \& R. A. M{\'e}ndez\inst{2}
          }

   \institute{Gemini Observatory, Southern Operations Center, c/o AURA, Casilla 603, La Serena, Chile\email{vkalari@gemini.edu}
   \and
Departamento de Astronom\'{\i}a, Universidad de Chile,
       Casilla 36-D Santiago, Chile
       \and
             Armagh Observatory, College Hill, Armagh, BT61\,9DG, UK        
             \and
             European Southern Observatory, Alonso de Cordova 3107, Casilla 19001, Santiago, Chile
\and
             Astrophysics Research Institute, Liverpool John Moores University, 146 Brownlow Hill, Liverpool L3 5RF, UK 
          }

   \date{}

 
  \abstract
   {The formation of the most massive stars in the Universe remains an unsolved problem. Are they able to form in relative isolation in a manner similar to the formation of solar-type stars, or do they necessarily require a clustered environment? In order to shed light on this important question, we study the origin of two very massive stars (VMS): the O2.5If*/WN6 star RFS7 ($\sim$100\,$M_{\odot}$), and the O3.5If* star RFS8 ($\sim$70\,$M_{\odot}$), found within $\approx$ 53 and 58\,pc respectively from the Galactic massive young cluster NGC\,3603, using {\it {\it Gaia}} data. RFS7 is found to exhibit motions resembling a runaway star from NGC 3603. This is now the most massive runaway star candidate known in the Milky Way. Although RFS8 also appears to move away from the cluster core, it has proper-motion values that appear inconsistent with being a runaway from NGC 3603 at the $3\sigma$ level (but with substantial uncertainties due to distance and age). Furthermore, no evidence for a bow-shock or a cluster was found surrounding RFS8 from available near-infrared photometry. In summary, whilst RFS7 is likely a runaway star from NGC\,3603, making it the first VMS runaway in the Milky Way, RFS8 is an extremely young ($\sim$2\,Myr) VMS, which might also be a runaway, but this would need to be established from future spectroscopic and astrometric observations, as well as precise distances. If RFS\,8 were still not meeting the criteria for being a runaway from NGC 3603 from such future data, this would have important ramifications for current theories of massive star formation, as well as the way the stellar initial mass function (IMF) is sampled.}


   \maketitle
%

\section{Introduction}

Whether nature is able to convert a cold molecular core into a single very massive star (VMS with $M_{\ast}>100M_{\odot}$) without
the contemporaneous production of a stellar cluster is strongly debated (Vink 2015). Fragmentation of the core would naturally ensue during overall gravitational collapse. Over-densities unstable to collapse are readily produced by turbulence. Feedback on the other hand
may or may not halt the continual fragmentation on progressively smaller scales. In the competitive accretion model
(Bonnell et al. 1998) unobstructed fragmentation creates a swarm of Jupiter mass objects which are unequally
fed from the wider clump material; the formation of a stellar cluster is unavoidable. Monolithic collapse
(Krumholz \& McKee 2008) envisions a self-gravitating and autonomous core that transforms into an individual
high-mass star in quite a similar way to that of a low-mass star, i.e. single and isolated. 

Strong observational support in favour of monolithic collapse would be provided by the detection of `isolated' very massive stars (Krumholz 2015). Yet increased spatial resolution observations invariably detect clusters around them (Stephens et al. 2017), or astrometric observations find that these stars are located in the field because they are runaways (e.g. de Wit et al. 2004; McSwain et al. 2007), strongly favouring the merger scenario. This highlights a problem: that it is difficult to prove observationally that a very massive star can form in isolation, although, not impossible. To do so convincingly, one requires the isolated O star to be (demonstrably) young, of the highest mass possible, with precise proper motions to trace its origin, and preferably nearby to allow imaging of low-mass siblings. Some candidates exist (e.g. Bressert et al. 2012; Oey et al. 2013). The most extreme of these candidates for isolated very massive star formation are found in 30 Doradus in the Large Magellanic Cloud ($d=51$\,kpc): for e.g. the very massive stars VFTS\,16, 72 (Lennon et al. 2018), and 682 (at $\sim$150\,$M_{\odot}$; Renzo et al. 2019) located at distances $\gtrsim$25\,pc from the starburst cluster R136 at the heart of 30 Doradus, where its twin VMS counterparts are located. Although recent {\it Gaia} and {\it HST} (Hubble Space Telescope) proper motions suggest that they maybe the most massive runaway stars found to date (at least for VFTS\,16 and 72), the issue is still under debate for VFTS\,682. 

Turning to the Milky Way, Roman-Lopes et al. (2016) presented evidence for two field very massive stars that may fulfil these criteria from
a study around the star-forming region NGC\,3603 ($d\sim$6-8\,kpc). They identified two VMS 
stars (RFS7=O2.5If*/WN6; RFS8=O3.5If*) with masses $\sim$100\,$M_{\odot}$  which are
situated $>$50\,pc from the centre of NGC\,3603\footnote{The centre of the star-forming region NGC\,3603 is the NGC\,3603 young cluster (NGC\,3603 YC), also designated as HD\,97950. Here on we refer to it as NGC\,3603 YC.}, with no {\it known} significant stellar over-densities within 2\,pc based on literature studies. They comprise $\sim$4\% of the known stellar content in the region earlier than O6 (Melena et al. 2008; Roman-Lopes et al. 2016). 
Interestingly, NGC3603 YC is considered the Galactic counterpart to R\,136, around 2.5 times smaller in massive star content. Both contain a halo of massive stars surrounding them, and Drew et al. (2019) found a series of canonical O star ejections for this cluster. 

In this letter, we investigate the origin of the two most massive stars in this sample (RFS\,7 and RFS\,8), whether they present {\it Gaia} DR2 motions consistent with a runaway status; or if they may have formed in-situ to help discriminate between isolated very massive star formation scenarios (Kurmholz 2015). This letter is organised as follows: in Section\,2 we present high-precision {\it {\it Gaia}} proper motion data to investigate if these stars originated from NGC\,3603. In Section\,3 we use archival near-infrared data to identify (if any) clusters surrounding these stars. Section\,4 presents a discussion of our results.  

\section{RFS7 and RFS8 as runaways}

\begin{table}
\caption{Spectral type and positions of RFS7 and RFS8} 
\resizebox{\columnwidth}{!}{%
\begin{tabular}{lrrrrrrr}
\hline
  \multicolumn{1}{c}{Star} &
  \multicolumn{1}{c}{SpT} &
  \multicolumn{1}{c}{$\alpha^1$ } &
  \multicolumn{1}{c}{$\delta^1$}  & \\
      \multicolumn{1}{c}{} &
  \multicolumn{1}{c}{} &
  \multicolumn{1}{c}{($\degr$)} &
  \multicolumn{1}{c}{($\degr$)}  \\
\hline\hline
  RFS7 & O2.5If*/WN6 & 168.8140 & $-$60.8549   \\
  (DR2\,5337456262235600512) & & & \\
  RFS8 & O3.5If* & 169.0526 & $-$61.7317   \\
  (DR2\,5337017247839725184) & & & \\
\hline\end{tabular}}
\tablefoot{($1$) All coordinates are in the 2015.5 epoch }
\end{table}

\subsection{Kinematic data}

All sources within 0.5$\arcmin$ of NGC\,3603 YC were extracted from the {\it Gaia} Release 2 (DR2) catalogue (Gaia Collaboration et al. 2018). Stars having proper motion and parallax errors greater than 0.05\,mas and 0.1\,mas respectively were discarded, and the filtering astrometric equations of Lindegren et al. (2018) were applied. In addition, the {\it Gaia} DR2 catalogue was cross-matched to the positions of RFS7 ({\it Gaia} DR2 5337456262235600512) and RFS8 ({\it Gaia} DR2 5337017247839725184) from Roman-Lopes et al. (2016) with a radius of 0.2$''$, and to the 38 massive stars in NGC\,3603 YC from Melena et al. (2008) with a radius of 0.3$''$. We compared the {\it Gaia} $BP$ and $G$ magnitudes to $BVI$ magnitudes of the respective parent catalogues using the transformation equations of Jordi et al. (2016). Stars offset by more than 0.5\,mag were discarded as discrepant matches. From tests performed by the {\it Gaia} collaboration, it is possible that close companions may result in spurious astrometric solutions. We examine 0.05$''$ {\it Hubble} images of the central cluster (at a nominal distance of 8\,kpc this angular resolution translates to 0.002\,pc, sufficient to separate resolve a 500\,A.U. binary in its components) for close companions, but find no visual companions. {\it Gaia} does not provide radial velocities for the isolated O stars of Roman-Lopes et al. (2016), but only proper motions in Right Ascension and Declination ($\mu_\alpha$, $\mu_\delta$), and parallaxes ($\pi$) with acceptable errors ($<$0.05\,mas\,yr$^{-1}$, $<$0.05\,mas). The summary of the observational data of the isolated O stars are given in Table\,1.  

\begin{figure}
\center
\includegraphics[width=78mm, height=60mm]{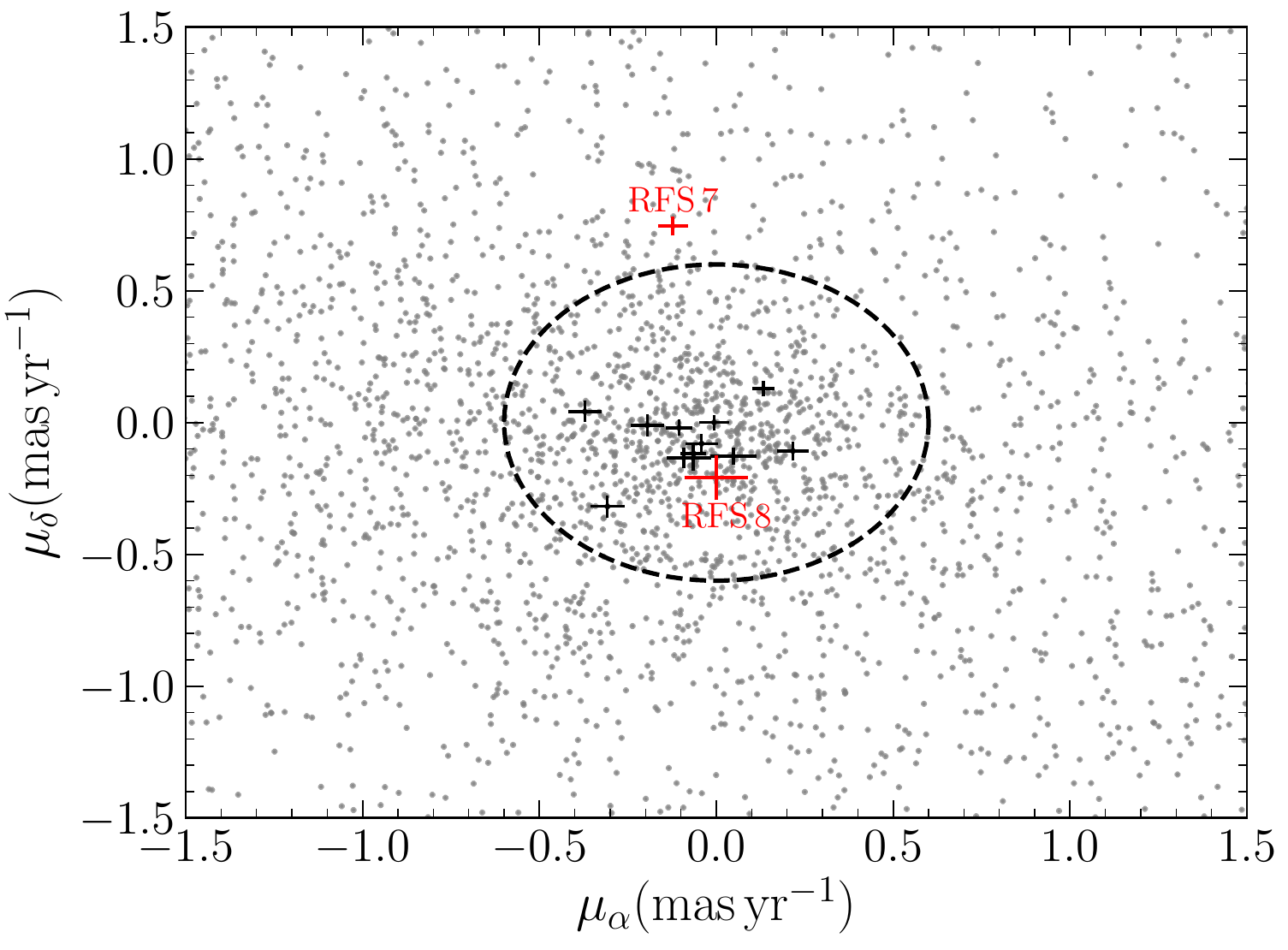}
\caption{Vector-point diagram of proper motions of in NGC\,3603, along with RFS\,7 and 8 normalised to the cluster centre. The black dots and the associated error bars are the proper motions of O stars meeting our quality criteria, and the red dots those of RFS\.7 and 8. The grey dots are the proper motions of all stars meeting our quality criteria within 5$\arcmin$ of NGC\,3603. The ellipse marks the boundary of the cluster motions (including for errors).}
\label{slope}
\end{figure}

The {\it Gaia} collaboration outline the parameters to gauge the precision (the formal uncertainties), reproducibility (checked using the parameters visibility periods, which are a group of observations separated from other such groups by at least four days), and accuracy and consistency of the astrometric solution (the reference unit weight error parameter, RUWE). We checked each of these for RFS7 and RFS8. Both stars have small formal uncertainties on their proper motions ($<$0.09\,mas) and parallaxes ($<$0.05\,mas), with sufficient total number of AL (along-scan) observations (NAL$>$200), to suggest the results are both precise and reproducible. In addition, we calculated the RUWE term{\footnote {This was calculated following the prescription in the {\it Gaia} public document GAIA-C3-TN-LU-LL-124-01 by Lindegren 2018}}. We find that this value is 1.01 for RFS7, and 1.237 for RFS8, well within the cut-off of 1.96 suggested. Overall, we find that the data of the isolated O stars are of sufficient quality to continue with our analysis. Additionally, we apply the same checks on all stars in NGC\,3603 YC, and the known massive stars of Melena et al. (2008). We are left with 30, and 13 stars from the two datasets with astrometric data meeting our quality criteria respectively.

\begin{figure*}
\center
\includegraphics[width=140mm, height=120mm]{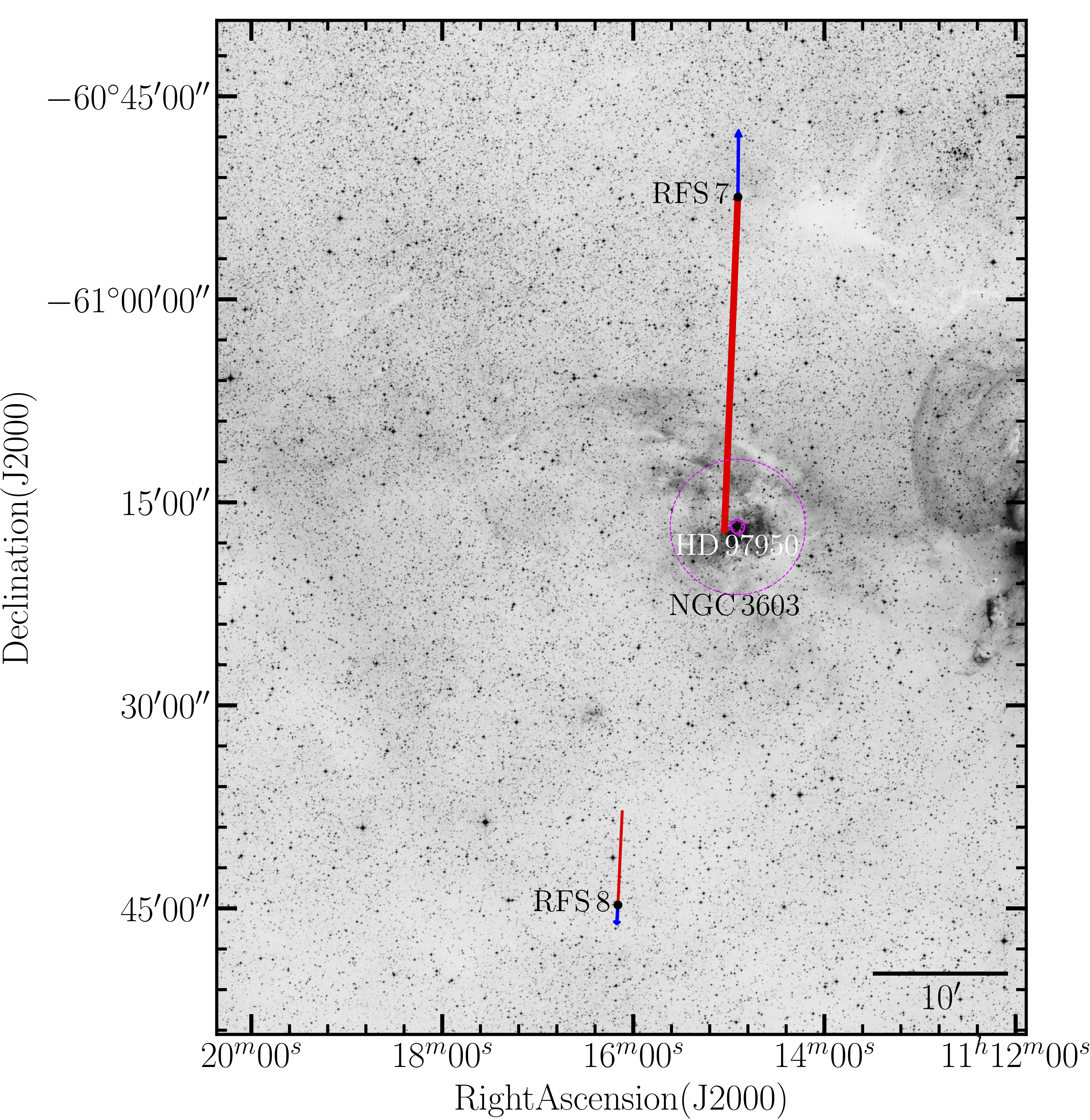}
\caption{{\it DSS} greyscale image centred on NGC\,3603. North is up and east is to the left. The blue arrows show the predicted motions relative to NGC\,3603 YC of RFS7 and RFS8, for a time-span of 0.4 Myr. The red lines in the opposite directions are proportional to the apparent age of the star, thus illustrating the site of potential origin. The magenta circles are centred on NGC\,3603 YC with a radii 0.5$\arcmin$ (solid line) and 5$\arcmin$ (dashed line) indicating the approximate boundaries of NGC\,3603 YC and NGC\,3603 respectively.}
\label{slope}
\end{figure*}

In the literature, the distances derived for NGC\,3603 span between 6--10\,kpc, with the most often quoted value of 7600\,pc derived from spectroscopic parallax of massive stars in the cluster centre (Melena et al. 2008), which is in agreement with CS line observations of N{\"u}rnberger et al. (2002) who found 7.7$\pm$1\,kpc. For our study, we assume that RFS\,7 and 8 lie at the same distance along the line of sight as the central cluster NGC\,3603 YC, and we adopt this distance to be 7.6\,kpc throughout. At this distance, the parallax value should be 0.13\,mas. However, the {\it Gaia} DR2 parallax suffers from systematic zero-point error of 0.03\,mas, with an additional error dependent on position varying by up to 0.1\,mas. Therefore {\it Gaia} DR2 does not provide precise constraints on the distance to NGC\,3603 YC, and RFS\,7 and 8, but can still be used to verify previous measures. In addition, we use the parallax to ascertain whether none of the two stars are significantly in the foreground/background relative to NGC\,3603 YC. To do so, we estimate the median distance to the central O stars in NGC\,3603 YC using the Bayesian inference method described in Bailer-Jones et al. (2018) adopting a length scale of 1350\,pc. We also shifted the parallax by the average zero-point error of 0.03\,mas (Lindegren et al. 2018), but we did not account for any potential position dependent variations. A median distance of 8525$\pm$1700\,pc is found for the O stars in the central cluster. Similarly, the distances to the isolated O stars RFS7 and RFS8 are similar to the central cluster, with a distance of 8.2 and 8.9\,kpc respectively. While a detailed discussion is beyond the scope of this paper, from the {\it Gaia} DR2 parallaxes we find that neither RFS\,7 or 8 appear to be significantly shifted along the sight with respect to NGC\,3603, and we assume they all fall at a distance of 7.6\,kpc in the rest of this letter.

\begin{table*}
\caption{{\it Gaia} DR2 proper motions, parallaxes and derived motions of RFS7 and RFS8}     
\begin{tabular}{lrrrrrrrr}
\hline
  \multicolumn{1}{c}{Star} &
  \multicolumn{1}{c}{$\mu_{\alpha}$} &
  \multicolumn{1}{c}{$\mu_{\delta}$} &
    \multicolumn{1}{c}{$\pi$} &
  \multicolumn{1}{c}{$\mu_{\rm relative}$} &    \multicolumn{1}{c}{$\mu_{l}$} &
  \multicolumn{1}{c}{$\mu_{b}$} &
  \multicolumn{1}{c}{$v_{\rm 2D}$} &
  \multicolumn{1}{c}{Pec. $v_{t}$} \\
    \multicolumn{1}{c}{} &
  \multicolumn{1}{c}{(mas\,yr$^{-1}$)} &
  \multicolumn{1}{c}{(mas\,yr$^{-1}$)} &
  \multicolumn{1}{c}{(mas)} &
  \multicolumn{1}{c}{(mas\,yr$^{-1}$)} &
  \multicolumn{1}{c}{(mas\,yr$^{-1}$)} &
  \multicolumn{1}{c}{(mas\,yr$^{-1}$)} &
  \multicolumn{1}{c}{(km\,s$^{-1}$)} &
  \multicolumn{1}{c}{(km\,s$^{-1}$)} \\
\hline\hline
  RFS7 &  $-$5.636$\pm$0.042 & 2.736$\pm$0.036 & 0.076$\pm$0.025 & 0.75$\pm$0.04 & $-$6.246$\pm$0.04	&	0.486$\pm$0.03 & 27.1 $\pm$1.5 & 33.3$\pm$1.9\\
  RFS8 &  $-$5.513$\pm$0.09 & 1.782$\pm$0.085 & $-$0.004$\pm$0.053 & 0.21$\pm$0.12 & $-$5.784$\pm$0.07	&	$-$0.336$\pm$0.08 & 7.6$\pm$2.7 & 10.2$\pm$4.5\\
  \hline\end{tabular}
\tablefoot{(1) All proper motions are in the 2015.5 reference frame. Note that the errors on the 2D and peculiar velocities do not account for the error on the distance. All motions were derived assuming $d$=7.6\,kpc. }
\end{table*}

\subsection{Relative proper motions}

The centre of NGC\,3603 YC in proper motion space was found to be $\mu_\alpha$=$-$5.5362$\pm$0.3\,mas\,yr$^{-1}$, $\mu_\delta$=1.9906$\pm$0.39\,mas\,yr$^{-1}$. This value was determined by fitting a Gaussian to the observed values, where the uncertainty is the standard deviation. We assume that these values best represent the central locus of the cluster in the $\mu_\alpha$--$\mu_\delta$ plane, and thereby converted all absolute proper motions relative to this value.

The resulting vector-point diagram of the proper motions is displayed in Fig.\,1. The central locus of the massive stars and those within 0.5$\arcmin$ of the cluster with high quality astrometric data is clearly identified, with surrounding ellipse identifying the cluster boundaries in proper motion space. Also plotted are the proper motions of all stars within 5$\arcmin$ of the cluster, and of RFS\,7 and RFS\,8. In the resulting diagram, RFS\,7 is identified as an outlier (at the 2$\sigma$ level). The proper motion errors ($<$0.07\,mas\,yr$^{-1}$), and quality checks on the data suggest that the {\it Gaia} astrometric results are of high quality with low errors. The star is also located in a region of low stellar density, with a projected distance of 53$\pm7$\,pc. In contrast, RFS8 appears within the locus of the central cluster in proper motion space, and has a combined relative proper motion of 0.21\,mas\,yr$^{-1}$. It displays high astrometric fidelity. However, it is located in a region of low stellar density around 59$\pm8$\,pc from the central cluster, prompting curiosity on its birth location. 

In Fig.\,2, we show the proper motion of RFS7 and 8 projected on the sky, where the direction and length of the arrows indicate the direction and value of the proper motion vector respectively. RFS7's proper motion indicates it is moving away from NGC\,3603 which is as expected for runaway stars, but even though the direction of RFS8's proper motion is that of an expected runaway, its magnitude is not. We compare the measured 2D velocity of each star with its predicted speed if ejected from the cluster on formation. This latter value is simply the projected distance over the isochronal age of the star (which is adopted from Roman-Lopes et al. 2016). For RFS7, we find the two values are nearly identical where the actual 2D velocity (27\,pc\,Myr$^{-1}$) $\approx$ predicted speed (31\,pc\,Myr$^{-1}$). The kinematic age of the star RFS7 is $\sim$1.8\,Myr. Given the age of the cluster (1--2\,Myr), this is consistent with a scenario of ejection during the initial stages of cluster formation. However, for RFS8, the predicted speed ($\sim$29\,pc\,Myr$^{-1}$) is nearly three times larger than the estimated 2D velocity (7\,pc\,Myr$^{-1}$). If hypothetically, we assume a smaller distance to RFS8 (7\,kpc), and uncertainty on the spectral type O4.5 supergiant (Roman-Lopes et al. 2016) we calculate a lower predicted velocity (15\,pc\,Myr$^{-1}$) more consistent to the 2D velocity. In this scenario, RFS8 cannot have been ejected during the formation of the central cluster (which formed no more than 2\,Myr ago), and must have been during an earlier star-formation event. Finally, we note that the absolute values of the {\it Gaia} proper motions of RFS8 would require significant revision in future releases to be consistent with a runaway origin.

Runaway stars typically exhibit high peculiar tangential velocities, typically greater than 30\,km\,s$^{-1}$ (e.g. Blaauw 1961; Stone 1979). Peculiar velocities are the stellar velocities corrected for Solar motion and Galactic rotation. We therefore estimate the peculiar tangential velocities. To do so, we adopted $U_{\odot}, V_{\odot}, W_{\odot}$= 10.0,11.0,7.2 km\,s$^{-1}$ respectively to correct for solar motion (McMillan \& Binney 2010), and adopted a flat rotation curve with the solar Galactocentric distance of 8.5\,kpc, with a rotation velocity of 220\,km\,s$^{-1}$ to correct for Galactic motion (Kerr \& Lynden-Bell 1986), suitable for the Galactocentric distance to NGC\,3603. This results in a peculiar tangential motion of 33.3\,km\,s$^{-1}$ for RFS7. The uncertainty here accounts only for the proper motion errors. But for RFS8, we derive a value of only 10.2\,km\,s$^{-1}$. Following the runaway criteria of Teztlaff et al. (2011) and Stone (1979), most runaways have peculiar tangential motions above 28\,km\,s$^{-1}$. From this, RFS7 is a {\it clear} runaway star (3$\sigma$ above the threshold), but RFS8 has velocities 3$\sigma$ less than the lower limit for runaways.

\subsection{Bow shocks}

We searched for bow shocks using mid-infrared imaging (Cutri et al. 2012) from the WISE (Wide-field Infrared Survey Explorer) space telescope. The WISE imaging has spatial angular resolutions in the 3.4, 4.6, 12, and 22\,$\mu$m bands (W1, W2, W3, W4) of 6.1$\arcsec$, 6.4$\arcsec$, 6.5$\arcsec$, \& 12.0$\arcsec$ respectively, which translates roughly to projected distances of 0.25--0.5\,pc, sufficient to identify most bow shocks around massive stars which are approximately few pc in size. 

Searching the WISE imaging, we found evidence for a bow shock around RFS\,7, but not RFS\,8 (see Fig.\,3). The combined {\it rgb} image, shows evidence for a bow shock shaped structure approximately 1\,pc from RFS7, with a length of $\sim$5\,pc. The structure is only visible at long wavelengths (W3 and W4 bands), but not in the short wavelength WISE imaging. This is because the W3 and W4 are excellent tracers of dust, but at W1 and W2 wavelengths, the stellar components dominate. Searching the imaging around RFS\,8, we find no evidence for a bow shock around RFS8. In fact, to the contrary we find that embedded dust emission saturates the star at long wavelengths, indicating the star is still embedded in some natal nebulosity which would not be the case if it was travelling at high velocities for a few Myr. This evidence points towards RFS7 being a runaway star, and RFS8 not. 

\begin{figure}
\center
\includegraphics[width=80mm, height=60mm]{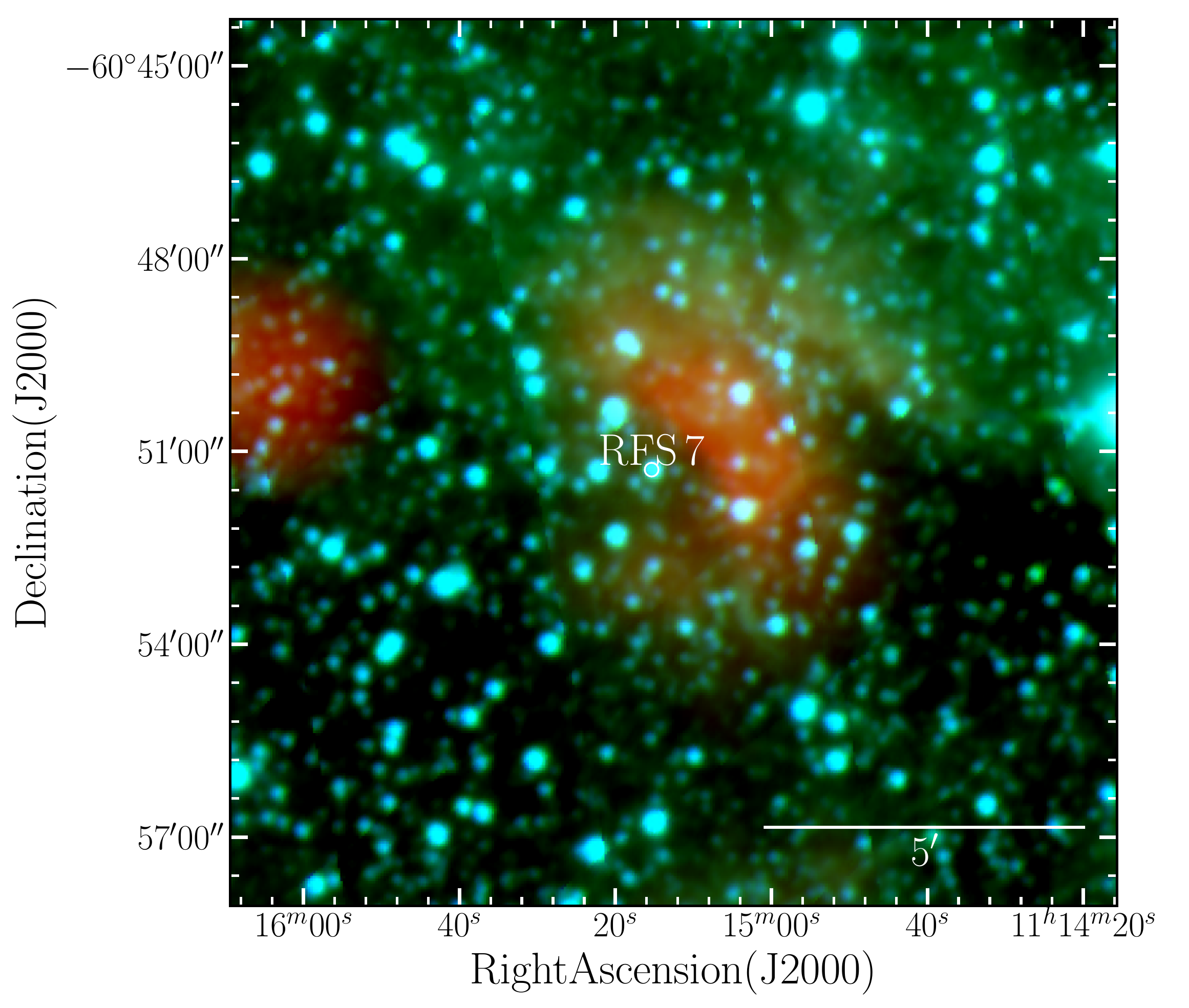}
\caption{WISE $r=22\mu$m/$g=4.6\mu$m/$b=3.4\mu$m image of RFS\,7. North is up and east is to the left. The bow shock is visible in the 22$\mu$m and is approximately 1\,pc away from the RFS7 towards the north west. The image resolution is around 12$\arcmin$ for the 22$\mu$m image, but 6$\arcmin$ for the short wavelength images. NGC\,3603 is located south east.}
\label{slope}
\end{figure}

Overall, what does our analysis imply? We find that the relative proper motions, kinematic ages, and peculiar tangential motions of RFS7 to be consistent with a runaway star from NGC\,3603 YC. Consistent with this picture, we find evidence for a bow shock in infrared imaging. For RFS8, the results give a somewhat mixed picture. The {\it Gaia} DR2 proper motion is not discrepant from the cluster centre, and it has a predicted speed larger than the observed 2D velocity (at $>3\sigma$ level), with a small peculiar tangential velocity ($\sim$3$\sigma$ from the lower limit for runaway stars).{\footnote {Based on this, recently Drew et al. (2019) ruled out the runaway criteria.}} No evidence for a bow shock is found. But, when considering the systematic uncertainties on the {\it Gaia} DR2 parallaxes ($\sim$0.03\,mas), and proper motions ($\sim$0.04\,mas\,yr$^{-1}$) and considering the direction of the proper motion away from the cluster, it is harder to conclusively rule out the runaway scenario. If we assume the {\it Gaia} DR2 distance along the line of sight ($\sim$8.9\,kpc) for RFS8, its peculiar tangential velocity is more consistent ($\sim$24\,km\,s$^{-1}$) with a runaway star. Similarly, if we assume a larger age ($\sim$3.5\,Myr; a spectral misclassification of around 1 subtype lower) the predicted speed is $\sim$15\,km\,s$^{-1}$ more in line with the proper motions. We therefore suggest that based on the current data RFS8 cannot be classified as a runaway (although the direction of its proper motion is consistent with a runaway), yet the uncertainties are not sufficiently low to completely rule out this possibility. Future {\it Gaia} releases, and ground-based spectroscopic observations are essential to do so. We also investigate whether the observed position of RFS8 can be explained as forming in-situ, either in a clustered mode, or possibly a site of isolated massive star formation.

\section{Did RFS8 formed in-situ?}

The other possibility to explain the location and the age of the massive O3.5If* star RFS8 being nearly 69\,pc from the nearest massive cluster is that the star formed in a small cluster, which has not yet been detected. To investigate if there is a cluster around RFS8, we obtain Two Micron All Sky Survey (2MASS) near-infrared $JHK$s imaging (Cutri et al. 2003). Currently, this is the deepest photometry (allowing for extinction) available in the vicinity of RFS8. We obtain photometry of a region of 5$\arcmin$ surrounding the cluster. The depth of the photometry reaches roughly 16.5\,mag in $K$s. Allowing for an extinction similar to RFS8 (of $A_V$=6.7), this reaches down to a spectral type of approximately mid-late A. Following the sequential sampling prescription of the stellar initial mass function (IMF) from Kroupa et al. (2001), we should have been able to detect few 10s of A type stars to allow for a formation of such a massive star. This should be sufficient to identify any young cluster ($<$3 Myr following the upper age limit of RFS8). Note that in the case of stochastic sampling of the IMF, there are no restrictions on the mass of the second most massive star in the cluster, i.e. there is still a possibility of finding a cluster, but with few BA stars. From the colour-colour and colour-magnitude diagrams of stars within 5$\arcmin$ of RFS8, having high-quality infrared photometry, we find no evidence for a cluster sequence, and the upper locus of stars in the colour-magnitude diagram have no clear spatial correlation. In addition, we plot the stellar density (Fig.\,5) of all point sources. From this, we find no evidence for any significant clustering around RFS8.

\section{Discussion}

\begin{figure}
\center
\includegraphics[width=78mm, height=60mm]{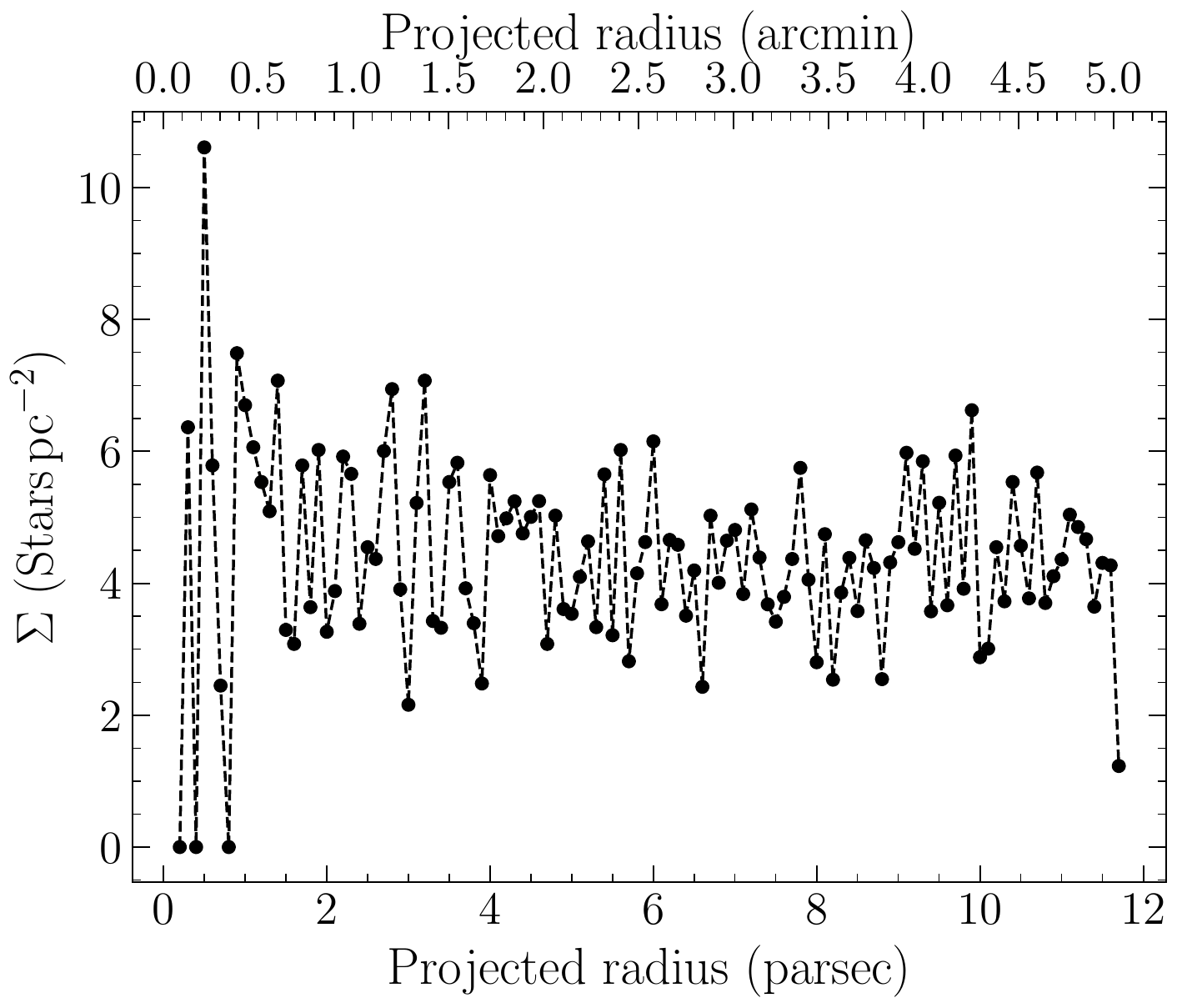}
\caption{Stellar density as a function of distance from RFS8 of all infrared 2MASS sources within 5$\arcmin$. The lower axis shows the radius in projected distance. }
\label{slope}
\end{figure}

We have shown that RFS7, a young ($\sim$1\,Myr), massive ($\sim$100\,$M_{\odot}$) O2.5If*/WN6 star is located $\approx$53\,pc from the centre of the nearest massive cluster, NGC\,3603 YC. Its mass is comparable to the most massive stars found within the cluster (Roman-Lopes et al. 2016). From {\it Gaia} astrometry, we find that RFS7 displays a proper motion outside the locus of cluster stars of NGC\,3603 YC, with a 2D projected velocity from NGC\,3603 YC of 27.1$\pm$1.5 km\,s$^{-1}$. It has a high tangential velocity ($\sim$33 km\,s$^{-1}$), and a possible bow shock located 1\,pc from the star. The motion suggests the star was ejected from the cluster 1-2\,Myr, consistent with the age of the cluster. All this indicates that RFS7 could likely be a runaway from NGC\,3603 YC, making it the most massive known runaway in our Galaxy, (and comparable to the behemoths found around the R\,136 cluster in the Large Magellanic Cloud). A key future study in this respect is understanding whether the runaway was dynamically ejected, which can also help us understand better the total mass and stellar content of the central cluster (to eject a $\sim$100\,$M_{\odot}$ star, the central cluster must have several binaries with total masses in excess of these masses, and produce further runaways, some which have been detected e.g. Gvaramadze et al. 2012). 

RFS8, another young ($\sim$2\,Myr), massive ($\sim$70\,$M_{\odot}$) O3.5If* star is located slightly further from NGC\,3603 YC, at 59\,pc. From accurate and precise {\it Gaia} data, we find that RFS8 displays a proper motion similar to the locus of central cluster stars (see Fig.\,1). It has an extremely low 2D projected velocity from the cluster, at 8 km\,s$^{-1}$. In contrast, its predicted speed if ejected from NGC\,3603 YC at birth would be around three times that value. Its peculiar tangential velocity is also lower than most runaway stars. Combined, we find no concrete evidence to suggest that RFS8 is a conventional runaway from NGC\,3603 YC, however current uncertainties do not preclude this possibility. We focused our attention on finding a cluster surrounding RFS8, but find that no significant clustered population of stars earlier than spectral type A were detected. Overall, we suggest that current data indicates RFS8 is not a runaway, yet future spectroscopic (age), further high precision astrometric (2D space motions, and tangential velocities) and radial velocities (3D space motions) are needed to completely rule out this possibility. We note RFS8 could not have been ejected from NGC\,3603 YC unless it had significantly higher space motions, and that it may also have been ejected in an earlier event in that case. If RFS8 is not a runaway, we note that in a sequentially sampled IMF, a few 10s of A type stars are expected to have formed to give rise to an early O type star which were not detected in current photometric surveys. Alternatively, some O stars are thought be ejected via a two-step process, where the less massive secondary is not traceable back to its origin (Pflamm-Altenburg \& Kroupa 2010), while dynamically ejected O stars may exhibit lower velocities around 10 km\,s$^{-1}$ under some conditions (Oh et al. 2015). We encourage further astrometric and spectroscopic study- to verify the {\it Gaia} proper motions, and to measure its radial velocity and total space motions. Combined with a deep infrared imaging (this is essential given the total extinction hovers around 6\,mag) to appreciate any previously undetected cluster, future studies may confirm the (un)isolated nature of RFS8. We also point out that if the star is in fact not a runaway, it indicates a more stochastic sampling of the IMF, given the lack of nearby massive stars around its potential formation site. Future studies are essential to probe further the origin of RFS8.

\begin{acknowledgements}
We thank the referee, Danny Lennon for helpful and constructive comments, and pointing out an error in the analyses. V.M.K. acknowledges funding from CONICYT Programa de Astronomia Fondo Gemini-Conicyt as GEMINI-CONICYT 2018 Research Fellow 32RF180005. This work is supported in part by the Gemini Observatory, which is operated by the Association of Universities for Research in Astronomy, Inc., on behalf of the international Gemini partnership of Argentina, Brazil, Canada, Chile, the Republic of Korea, and the United States of America. NB gratefully acknowledges financial support from the Royal Society (University Research Fellowship) and the European Research Council (ERC-CoG-646928, Multi-Pop). RAM acknowledges support from CONICYT/FONDECYT grant 1170854 and from CONICYT project Basal AFB-170002. This work has made use of SIMBAD and Vizier database, operated at CDS, Strasbourg, France. This work has made use of data from the European Space Agency (ESA) mission
{\it Gaia} (\url{https://www.cosmos.esa.int/Gaia}), processed by the {\it Gaia}
Data Processing and Analysis Consortium (DPAC,
\url{https://www.cosmos.esa.int/web/Gaia/dpac/consortium}). Funding for the DPAC
has been provided by national institutions, in particular the institutions
participating in the {\it Gaia} Multilateral Agreement. This publication makes use of data products from the Two Micron All Sky Survey, which is a joint project of the University of Massachusetts and the Infrared Processing and Analysis Center/California Institute of Technology, funded by the National Aeronautics and Space Administration and the National Science Foundation. This publication makes use of data products from the Wide-field Infrared Survey Explorer, which is a joint project of the University of California, Los Angeles, and the Jet Propulsion Laboratory/California Institute of Technology, funded by the National Aeronautics and Space Administration. 
\end{acknowledgements}


\end{document}